\documentclass[11pt]{amsart}

\usepackage[utf8]{inputenc}
\usepackage{amsmath,cite,url,amssymb,amsthm,amsfonts}
\usepackage{microtype}
\usepackage{graphicx, placeins,todonotes}
\usepackage{tikz}
\usetikzlibrary{calc,shapes,positioning,arrows}
\usepackage{color}
\usepackage[numbers,square]{natbib} 
\usepackage{hyperref}

\begin{document}

\author[A. Marafioti]{Andr\'es Marafioti{$^\dag$$^\ddag$}{}}
\address{$^\dag$ Acoustics Research Institute,
Austrian Academy of Sciences,
Wohllebengasse 12--14,
A-1040 Vienna,
Austria}
\thanks{This work has been supported by Austrian Science Fund (FWF) project MERLIN (Modern methods for the restoration of lost information in digital signals;I 3067-N30). We gratefully acknowledge the support of NVIDIA Corporation with the donation of the Titan X Pascal GPU used for this research.\\
\textit{This is the Author's Accepted Manuscript version of work presented at the 146th Convention of the Audio Engineering Society. It is licensed under the terms of the \href{https://creativecommons.org/licenses/by/4.0/}{Creative Commons Attribution 4.0 International License}, which
permits unrestricted use, distribution, and reproduction in any medium, provided the original author and source are credited. The published version is available at:} \url{https://www.aes.org/e-lib/browse.cfm?elib=20303}}
\email{amarafioti@kfs.oeaw.ac.at}
\author[N. Holighaus]{Nicki Holighaus{$^\dag$}{}}
\email{nicki.holighaus@oeaw.ac.at}
\author[P. Majdak]{Piotr Majdak{$^\dag$}{}}
\email{piotr.majdak@oeaw.ac.at}
\author[N. Perraudin]{Nathana\"el Perraudin{$^\ast$}{}}
\address{$^\ast$ Swiss Data Science Center, ETH Z\"urich, Universit\"atstrasse 25, 8006 Z\"urich}
\email{nathanael.perraudin@sdsc.ethz.ch}

\title[Audio inpainting of music by means of neural networks]
      {Audio inpainting of music by means of neural networks}


\begin{abstract}
We studied the ability of deep neural networks (DNNs) to restore missing audio content based on its context, a process usually referred to as audio inpainting. We focused on gaps in the range of tens of milliseconds. The proposed DNN structure was trained on audio signals containing music and musical instruments, separately, with 64-ms long gaps. The input to the DNN was the context, i.e., the signal surrounding the gap, transformed into time-frequency (TF) coefficients. Our results were compared to those obtained from a reference method based on linear predictive coding (LPC). For music, our DNN significantly outperformed the reference method, demonstrating a generally good usability of the proposed DNN structure for inpainting complex audio signals like music.
\end{abstract}

\maketitle

\renewcommand{\thefootnote}{\fnsymbol{footnote}}
\footnotetext[3]{Corresponding author}
\renewcommand*{\thefootnote}{\arabic{footnote}}

\markright{}

\section{Introduction}\label{sec:introduction}

Audio processing tasks often encounter locally degraded or even lost information. Some common examples are corrupted audio files, lost information in audio transmission, and audio signals locally contaminated by noise. Audio inpainting deals with the reconstruction of lost information in audio. Reconstruction is usually aimed at providing coherent and meaningful information while preventing audible artifacts so that the listener remains unaware of any occurred problem. Successful algorithms are limited to deal with a specific class of audio signals, they focus on a specific duration of the problematic signal parts, and exploit a-priori information about the problem. In this work, we explore a new machine-learning algorithm with respect to the reconstruction of audio signals. From all possible classes of audio signals, we limit the reconstruction to either music, or individual musical instruments. We focus on the duration of the problematic signal parts, i.e., gaps, being in the range of tens of milliseconds. Further, we exploit the available audio information surrounding the gap, i.e., the context.

The proposed algorithm is based on an unsupervised feature-learning algorithm driven by context-based sample prediction. It relies on a neural network with convolutional and fully-connected layers trained to generate sounds being conditioned on its context. Such an approach was first introduced for images~\cite{Pathak2016} where the terminology of the context encoder (CE) was coined as an analogy to auto encoders~\cite{Goodfellow-et-al-2016}. Hence, we treat our algorithm as an audio-inpainting context encoder.

An application of the convolutional network directly on the time-domain audio signals requires extremely large training datasets for good results~\cite{Pons2017}. In order to reduce the size of required datasets, our CE works on chunks of time-frequency (TF) coefficients calculated from the time-domain audio signal. Trained with the TF coefficients, our CE is aimed to recover the lost TF coefficients within the gap based on provided TF coefficients of the gap’s surroundings. The TF coefficients were obtained from an invertible representation, namely, a redundant short-time Fourier transform (STFT)\cite{po76,gr01}, in order to allow a robust synthesis of the reconstructed time-domain signal based on the network output. Our CE reconstructs the magnitude coefficients only, which are then applied to a phase-reconstruction algorithm in order to obtain complex valued TF coefficients required for synthesizing the time-domain signal. From accurate TF magnitude information, phaseless reconstruction methods are known to provide perceptually close, often indiscernible, reconstruction despite the resulting time-domain waveforms usually being rather different.

Two network structure we considered reconstructs the magnitude coefficients only, which are then applied to a phase-reconstruction algorithm in order to obtain complex-valued TF coefficients required for synthesising the time-domain signal. From accurate TF magnitude information, phaseless reconstruction methods such as~\cite{griffin1984signal,perraudin2013fast,pruuvsa2017noniterative} are known to provide perceptually close, often indiscernable, reconstruction despite the resulting time-domain waveforms usually being rather different. 

\subsection{Related deep-learning techniques}\label{sec:relatedAlsoAI}

Deep learning excels in classification, regression, and anomaly detection tasks~\cite{Goodfellow-et-al-2016} and has been recently successfully applied to audio~\cite{schlueter2017_phd}. Deep learning has also shown good results in generative modeling with techniques such as variational auto-encoders\cite{KingmaW13} and generative adversarial networks~\cite{GoodfellowGAN2014}. Unfortunately, for audio synthesis only the latter has been studied and to limited success~\cite{Donahue2018}. The state-of-the-art audio signal synthesis require sophisticated networks, \cite{Mehri2016,Wavenet2016} in order to obtain meaningful results. While these approaches directly predict audio samples based on the preceding samples, in the field of text-to-speech synthesizing audio in domains other than time such as spectrograms~\cite{tacotron}, and mel-spectrograms~\cite{Tacotron2_2017} have been proposed. In the field of speech transmission, DNNs have been used to achieve packet 
loss concealment \cite{Lee2016PLC}.

The synthesis of \textit{musical} audio signals using deep learning, however, is even more challenging~\cite{dieleman2018challenge}. A music signal is comprised of complex sequences ranging from short-term structures (any periodicity in the waveform) to long-term structures (like figures, motifs, or sections). In order to simplify the problem brought by long-range dependencies, music synthesis in multiple steps has been proposed including an intermediate symbolic representation like MIDI sequences~\cite{Boulanger2012ModelingTD}, and features of a parametric vocoder~\cite{Blaauw2017}.

While all of these contributions can provide insights on the design of a neural network for audio synthesis, none of them addresses the specific setting of audio inpainting where some audio information has been lost, but some of its context is known. In particular in this contribution, we explored the setting in which the the audio information surrounding a missing gap is available.

\subsection{Related audio-inpainting algorithms}\label{sec:relatedAI}

Audio inpainting techniques for time-domain data loss compensation can be roughly divided into two categories: (a) Methods that attempt to recover precisely the lost data relying only on very local information in the direct vicinity of the corruptions. They are usually designed for reconstructions of gap with a duration of less than $10$~ms and also work well in the presence of randomly lost audio samples. (b) Methods that aim at providing a perceptually pleasing occlusion of the corruption, i.e., the corruption should not be annoying, or in the best case undetectable, for a human listener. The proposed restorations may still differ from the lost content. Such approaches are often based on self similarity, require a more global analysis of the degraded audio signal, and rely heavily on repetitive structures in audio data. They often cope with data loss beyond hundreds or even thousands of milliseconds.

In either of these categories, successful methods often depend on analyzing TF features of the audio signal instead of the time-domain signal itself. In the first category, we highlight two approaches based on the assumption that audio data is expected to possess approximately sparse time-frequency representations: In particular, variations of orthogonal matching pursuit (OMP) with time-frequency disctionaries~\cite{Adler2011, Adler2012,toumi:hal-01680669}, as well as (structured) $\ell^1$-regularization~\cite{Siedenburg2013AudioIW,Lieb2018}. They have been successfully applied to reconstruct gaps of up to 10-ms duration, however it is evident that neither of these methods is competitive when treating longer gaps. In the second class of methods, a method for packet loss concealment based on MFCC feature similarity and explicitly targeting a perceptually plausible restoration was proposed~\cite{Bahat2015}. Similarly, exemplar-based inpainting was performed on a scale of seconds based on a graph encoding 
spectro-temporal similarities within an audio signal~\cite{Perraudin2016}. In both studies, gap durations where beyond several hundreds of milliseconds and their reconstruction needed to be evaluated in psychoacoustic experiments.

In our contribution, we target gap durations of tens of milliseconds, a scale where the non-stationary characteristic of audio already becomes important, but a sample-by-sample extrapolation of the missing information from the context data still seems to be realistic. The methods from category (a) can not be expected to provide good results for such long gaps, and the methods from category (b) do not aim at reconstructions on the sample-by-sample level. Interestingly, the combination of that gap durations and that level of signal reconstructions do not seem to have received much attention yet. 

For simple sounds like those of musical instruments, linear prediction coding (LPC) can be applied. Assuming a modeling of the sound as an acoustic source filtered by a pole filter, extrapolation based on linear prediction has been shown to work well for gaps in the range of $5$ to $100$ milliseconds, e.g., \cite{Etter1996,kauppinen2002audio}. 
Although the performance of linear prediction relies heavily on the underlying stationarity assumption to be fulfilled, it seems to be the only competitive, established method in the considered scenario. 

Deep-learning techniques on the other hand, some of which we study here, promise a more generalized signal representation and therefore better results, whenever the lost data cannot be predicted by linear filtering. However, a link of deep-learning techniques with audio inpainting seems to be missing until now. 

\section{Context Encoder}\label{sec:ContextEncoder}

We consider the audio signal $s\in\mathbb{R}^L$, containing $L$ samples of audio. The central $L_g$ samples of $s$ represent the gap $s_g$, while the remaining $L_c$ samples on each side of the gap from the context. We distinguish between $s_b$ and $s_a$, which is the context signals before and after the gap, respectively. 

The architecture of our network is an encoder-decoder pipeline fed with the context information. Instead of passing the time-domain signals $s_b$ and $s_a$ directly to the network, the audio signal is processed to obtain TF coefficients, $S_b$, and $S_a$, which is the input to the encoder. The TF representation is propagated through the encode and decoder, both trained to predict TF coefficients representing the gap, $S'_g$. That output of the decoder is then post-processed in order to synthesize a reconstruction in the time domain, $s'$.

The network structure is comprised of standard, widely-used building blocks, i.e., convolutional and fully-connected layers, and rectified linear units (ReLUs). It is inspired by the context encoder for image restoration~\cite{Pathak2016}.\footnote{Before fixing the network structure described in the remainder of this section, we eperimented with different architectures, depths, and kernel shapes, out of which the current structure showed the most promise. Additionally, we also considered dropout~\cite{srivastava2014dropout} and skip connections, discarding them after not achieving any notable improvements.} Moreover, for the training, we do not use an adversarial discriminator, but optimize an adapted $\ell^2$-based loss.

The network was implemented in Tensorflow~\cite{tensorflow2015-whitepaper}. For the training, we applied the stochastic gradient descent solver ADAM~\cite{Kingma2014}. Our software, along with instructive examples, is available to the public.\footnote{www.github.com/andimarafioti/audioContextEncoder}

\subsection{Pre-processing stage}
In the pre-processing stage, STFTs are applied on the context $s_b$ and $s_a$ yielding $S_b$ and $S_a$, respectively. They are then split into real and imaginary parts, resulting in four channels $S_b^{Re},S_b^{Im}, S_a^{Re},S_a^{Im}$. The STFT is determined by the window $g$, hop size $a$ and the number of frequency channels $M$. In our study, $g$ was an appropriately normalized Hann window of length $M$ and $a$ was $M/4$, enabling perfect reconstruction by an inverse STFT with the same parameters and window. In order to obtain coefficients without artifacts even at the context borders, $s_b$ and $s_a$ were extended with zeros to the length of $L_c+3a$ towards the gap.

\vspace{4pt}
\subsection{Encoder}\label{par:encoder}

\begin{figure*}[!th]
\begin{center}
	\includegraphics[width=\textwidth,height=.25\textheight]{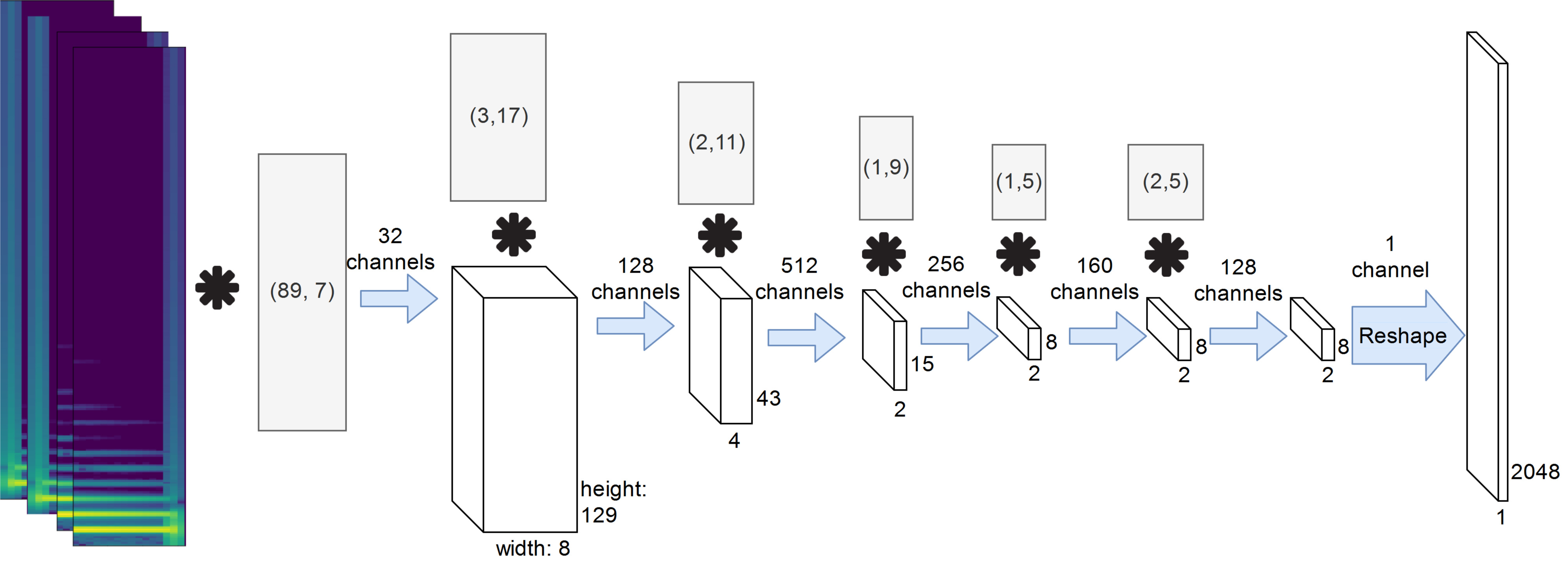}
	\caption{The encoder is a convolutional network with six layers followed by reshaping. The four channel time-frequency input is transformed into an encoding of size $2048$. Gray rectangles represent the convolution filters with size expressed as (height, width). White cubes represent the signal.}
	\label{fig:encoder}
\end{center}
\end{figure*}

The encoder is a convolutional neural network. The inputs $S_b^{Re},S_b^{Im}, S_a^{Re},S_a^{Im}$ of the context information are treated as separate channels, thus, the network is required to learn how the channels interact and how to mix them. Similar to \cite{Pathak2016}, all layers are convolutional and sequentially connected via ReLUs~\cite{Ramachandran2017}, after which batch normalization~\cite{Ioffe2015} is applied. The
resulting encoder architecture is shown in Figure \ref{fig:encoder}, for $M=512$.

Note that because the encoder is comprised of only convolutional layers, the information can not reliably propagate from one end of the feature map to another. This is a consequence of convolutional layers connecting all the feature maps together, but never directly connecting all locations within a specific feature map~\cite{Pathak2016}. 

\vspace{4pt}
\subsection{Decoder}

\begin{figure*}[!th]
\begin{center}
	\includegraphics[width=\textwidth,height=.2\textheight]{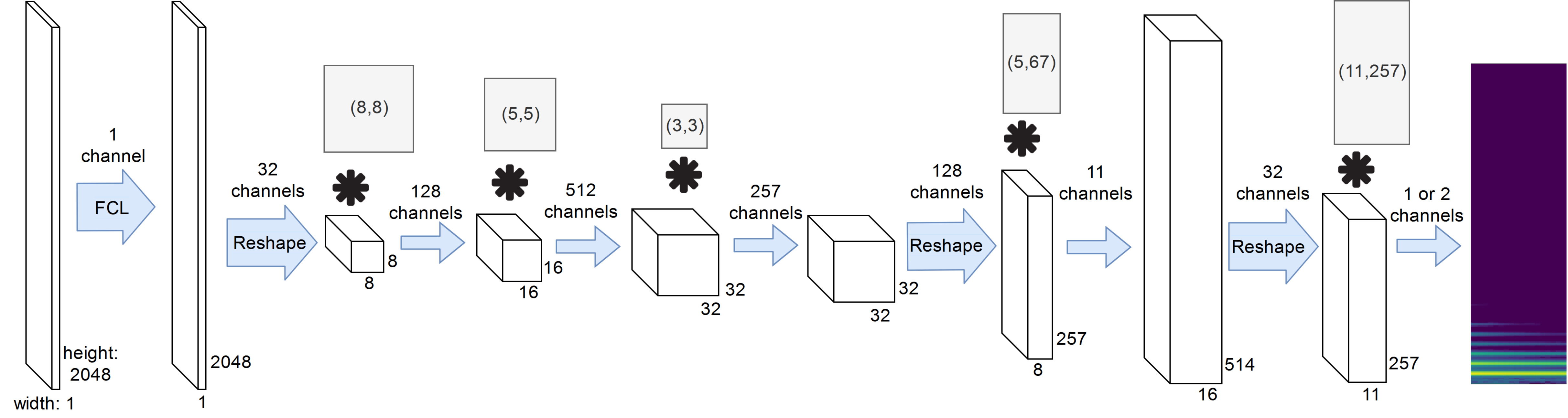}
	\caption{The decoder generates magnitude time-frequency coefficients from the encoding produced by the encoder, Fig. \ref{fig:encoder}. It consists of a fully connected layer (FCL) and five deconvolution layers, with reshaping after the fully-connected, as well as the third and fourth deconvolution layers. All other conventions as in Figure \ref{fig:encoder}.}
	\label{fig:decoder}
	\end{center}
\end{figure*}

The decoder begins with a fully connected layer (FCL) with a ReLU nonlinearity in order to spread the encoder information among the channels.\footnote{Fully connected layers are computationally very expensive; in our case it contains 38\% of all the parameters of the network. In \cite{Pathak2016}, this issue was addressed by using a 'channel-wise fully connected layer'. We tested that approach but obtained consistently worse results.} Similar to \cite{Pathak2016}, all the subsequent layers are (de-)convolutional and, as for the encoder, connected by ReLUs with batch normalization. The network architecture is shown in Figure \ref{fig:decoder}, for $M=512$ and a gap size of $1024$ samples, i.e., every output channel is of size $257\times 11$. 

The final layer depends on the network. For the complex network, the final layer has two outputs, corresponding to the real and imaginary part of the complex-valued TF coefficients. For the magnitude network, the final layer has a single output for the magnitude TF coefficients. We denote the output TF coefficients as $S'_g$.

\subsection{Post-processing stage}

The aim of the post-processing stage is to synthesize the reconstructed audio signal in the time domain. To this end, the reconstructed gap TF coefficients from the decoder, $S'_g$, are inserted between the TF coefficients of the context, $S_b$ and $S_a$. However, the gap between $S_b$ and $S_a$ is smaller then the size of $S'_g$' because the context coefficients were calculate from zero-padded time-domain signals. The coefficients at the context border represent the zero-padded information and are discarded for synthesis, forming $S'_b$ and $S'_a$. The sequence $S' = (S'_b, S'_g, S'_a)$ then has the same size as $S$. 
By performing the insertion directly in the time-frequency domain, we prevent transitional artifacts between the context and the gap, since synthesis by the inverse STFT introduces an inherent cross-fading. 

The decoder output represents the magnitudes of the TF coefficients and the missing phase information needs to be estimated separately. First, the phase gradient heap integration algorithm proposed in~\cite{ltfatnote043} was applied to the magnitude coefficients produced by the decoder in order to obtain an initial estimation of the TF phase. Then, this estimation was refined by applying 100 iterations of the fast Griffin-Lim algorithm~\cite{griffin1984signal, perraudin2013fast} implemented in the Phase Retrieval Toolbox Library~\cite{ltfatnote045}.\footnote{The combination of these two algorithms provided consistently better results than separate application of either.} The resulting complex-valued TF coefficients $S'_g$ were then transformed into a time-domain signal $s'$ by inverse STFT. 

\subsection{Loss Function}

The network training is based on the minimization of the total loss of the reconstruction. Generally, we computed the reconstruction loss by comparing the original gap TF coefficients $S_g$ with the reconstructed gap TF coefficients $S'_g$. The comparison is then done on the basis of the squared $\ell^2$-norm of the difference between $S_g$ and $S'_g$ as it is customary for this type of network~\cite{ZhaoLoss2017}, commonly known as mean squared error (MSE). The absolute MSE depends, however, on the total energy of $S_g$, clearly putting more weight on signals containing more energy. 

In contrast, the comparison can be done by using the \textit{normalized} mean squared error (NMSE), 

\begin{equation}
\mathbf{NMSE}(S_g,S'_g) = \frac{\Vert S_g - S'_g \Vert ^2}{\Vert S_g \Vert ^2}
\label{eq:MSEformula}
\end{equation}

While theoretically invariant to amplitude changes, this measure strongly amplifies 
small errors when the energy of $S_g$ is small. In practice, however, very minor deviations from $S_g$ are insignificant regardless of the content of $S_g$, which is not reflected in NMSE. 

Therefore, we propose to use a weighted mix between MSE and NMSE for the calculation of the loss function. This leads to 
\begin{equation}
\mathbf{F}(S_g,S'_g) = \frac{\|S_g - S'_g\|^2}{c^{-1} + \|S_g\|^{2}},
\label{rLoss}
\end{equation}
where the constant $c>0$ controls the incorporated compensation for small amplitude. In our experiments, $c=5$ yielded good results. 

Finally, as proposed in \cite{Krogh92asimple}, the total loss is the sum of the loss function and a regularization term controlling the trainable weights in terms of their $\ell^2$-norm:

\begin{equation}
\mathbf{T} = F(S_g,S'_g) + \frac{\lambda }{2} \sum_i w_i^2,
\label{tLoss}
\end{equation}

with $w_i$ being weights of the network and $\lambda$ being the regularization parameter, here set to $0.01$.

\section{Evaluation}

The main objectives of the evaluation were to investigate the general ability of the networks to adapt to the considered class of audio signals, as well as how they compare to the reference method, i.e., LPC-based extrapolation as proposed in~\cite{kauppinen2002audio}. Additionally, we considered the effects of changing the gap duration. 

To this end, we considered two classes of audio signals: instrument sounds, and music. For instruments and music, the network was trained on the targeted signal class, with an assumed gap size of $64$~ms. Reconstruction was evaluated on the trained signal class and pure tones for $64$~ms gaps. Additionally, the network was evaluated for $48$~ms gaps. Reconstruction quality was evaluated by means of signal-to-noise ratios (SNR) applied to the magnitude spectrograms, to accommodate for perceptually irrelevant phase changes. Further, all results were compared to the reconstruction based on the reference method.

\subsection{Parameters}
\label{sec:parameters}

The sampling rate was $16$~kHz. We considered audio segments with a duration of $320$~ms, which corresponds to $L=5120$ samples. Each segment was separated in a gap of $64$~ms (or $48$~ms, corresponding to $L_g=1024$ or $L_g=768$ samples) of the central part of a segment and the context of twice of $128$~ms (or $136$~ms), corresponding to $L_c=2048$ (or $L_c=2176$) samples. Both, the size of the window $g$ and the number of frequency channels $M$ were fixed to $512$~samples. Consequently, $a$ was $128$~samples, and the input to the encoder was $S_b, S_a\in\mathbb{C}^{257\times 16}$. 

\subsection{Datasets}

\label{Dataset}

The dataset representing musical instruments was derived from the NSynth dataset~\cite{nsynth2017}. NSynth is an audio dataset containing 305,979 musical notes from 1,006 instruments, each with a unique pitch, timbre, and envelope. Each example is four seconds long, monophonic, and sampled at $16$~kHz.

The dataset representing music was derived from the free music archive (FMA)~\cite{fma2017}. The FMA is an open and easily accessible dataset, usually used for evaluating tasks in musical information retrieval (MIR). We used the small version of the FMA comprised of 8,000 $30$-s segments of songs with eight balanced genres sampled at $44.1$~kHz. We resampled each segment to the sampling rate of $16$~kHz.

The original segments in the two datasets were processed to fit the evaluation parameters. First, for each example the silence was removed. Second, from each example, pieces of the duration of $320$~ms were copied, starting with the first segment at the beginning of a segment, continuing with further segments with a shift of $32$~ms. Thus, each example yielded multiple overlapping segments $s$. Note that for a gap of $64$~ms, the segment can be considered as a 3-tuple by labeling the first $128$~ms as the context before the gap $s_b$, the subsequent $64$~ms as the gap $s_g$, and the last $128$~ms as the context after the gap $s_a$. Finally, all segments with RMS smaller than ten to the negative four in $s_g$ were discarded.

The datasets were split into training, validation, and testing sets. For the instruments, we used the splitting proposed by~\cite{nsynth2017}. The music dataset, was split into approximately 70\%, 20\% and 10\%, respectively. The statistics of the resulting sets are presented in Table~\ref{dataset_count}.

\begin{table}[!th]
\centering
\begin{tabular}{@{}lll@{}}
\hline
                  & Count             & Percentage \\ \hline
Instruments training      & 19.4M        & 94.1               \\
Instruments validation & 0.9M           & 4.4               \\
Instruments testing    & 0.3M           & 1.5               \\ 
Music training         & 5.2M         & 70.0               \\ 
Music validation       & 1.5M         & 20.0               \\ 
Music testing          & 0.7M           & 10.0               \\ \hline \\
\end{tabular}
\caption{Datasets used in the evaluation.}
\label{dataset_count}
\vspace{-1em}
\end{table}

\subsection{Training}

The network was trained separately for the instrument and music dataset. This resulted in two trained networks. 

Each training started with the learning rate of $10^{-3}$. The reconstructed phase was not considered in the training. Every 2000 steps, the training progress was monitored. To this end, the network's output was calculated for the music validation dataset and the NMSE was calculated between the predicted and the actual TF coefficients of the gap. When converging, which usually happened after approximately 600k steps, the learning rate was reduced to $10^{-4}$ and the training was continued by additional 200k steps.\footnote{We also considered training on the instrument training dataset (800k steps) followed by a refinement with the music training dataset (300k steps). While it did not show substantial differences to the training performed on music only, a pre-trained network on instruments with a subsequent refinement to music may be an option in applications addressing a specific music genre.}

\subsection{Evaluation metrics}

For the evaluation of our results, in general, we calculated the SNR (in dB) 
\begin{equation}
\mathbf{SNR}(x,x') = 10 ~ \log \frac{\Vert x \Vert ^2}{\Vert  x - x' \Vert ^2}
\label{eq:SNRformula}
\end{equation}
for each segment of a testing dataset separately, and then averaged all these SNR across all segments of that testing dataset.

For the evaluation in the TF domain, we calculated $\mathbf{SNR}(|S_g|,|S_{rg}|)$, where $S_{rg}$ represents the central $5$ frames of the STFT computed from the restored signal $s'$ and thus represents the restoration of the gap. In other words, we compute the SNR between the spectrograms of the original signal and the restored signal, but only in the region of the gap. We refer to the average of this metric (across all segments of a testing dataset) to as $\text{SNR}_{\text{MS}}$, where MS references to magnitude spectrogram. Note that  $\text{SNR}_{\text{MS}}$ is directly related to the logarithmic inverse of the spectral convergence proposed in~\cite{sturmel2011signal}.

\FloatBarrier

\subsection{Reference method}

We compared our results to those obtained with a reference method based on LPC~\cite{Tremain1982}. LPC is particularly widely used for the processing of speech \cite{rajman2007speech}, but also frequently for extrapolation of audio signals~\cite{kauppinen2001method, kauppinen2002audio}.

For the implementation, we followed~\cite{kauppinen2002audio}, especially~\cite[Section 5.3]{kauppinen2002audio}. 
In detail, the context signals $s_b$ and $s_a$ were extrapolated onto the gap $s_g$ by computing their impulse responses and using them as prediction filters for a classical linear predictor. The impulse responses were obtained using Burg's Method~\cite{Burg1967} and were fixed to have 1000 coefficients according to the suggestions from~\cite{kauppinen2001method} and \cite{kauppinen2002reconstruction}. Their duration was the same as that for our context encoder in order to provide the same amount of context information. The two extrapolations were mixed with the squared-cosine weighting function. Our implementation of the LPC extrapolation is available online\footnote{www.github.com/andimarafioti/audioContextEncoder}.

Finally, we evaluate the results produced by the reference method in the same way as we evaluate the results produced by the networks: predictions were calculated for each segment from both testing datasets and the $\text{SNR}_{\text{TD}}$ as well as the $\text{SNR}_{\text{MS}}$ were calculated on the predictions. 

\section{Results and discussion}\label{sec:compare}

\subsection{Comparison to the reference method}

Table \ref{tab:SpectralDiv64} provides the $\text{SNR}_{\text{MS}}$ for the LPC-based reference reconstruction method. When tested on music, on average, the magnitude network outperformed the LPC-based method by 1.4~dB. When tested on instruments, the magnitude network underperformed by 8.6~dB. 

\begin{table}[!th]
	\centering
	\begin{tabular}{lcccccc}
		\hline
		& \multicolumn{2}{c}{Music} & \multicolumn{2}{c}{Instruments} \\
		& Mag   & LPC   & Mag    & LPC   \\ \hline
		Mean & 7.7    & 6.3   & 22.0   & 30.6  \\
		Std  & 4.3   & 5.1   & 10.4   & 18.9  \\ \hline \\
	\end{tabular}
	\caption{$\text{SNR}_{\text{MS}}$ (in dB) of reconstructions of 64~ms gaps for the magnitude networks and the LPC-based method.}
	\label{tab:SpectralDiv64}
	\vspace{-1em}
\end{table}

\begin{figure}[!th]
\begin{center}
\includegraphics[width=0.45\textwidth,height=.18\textheight]{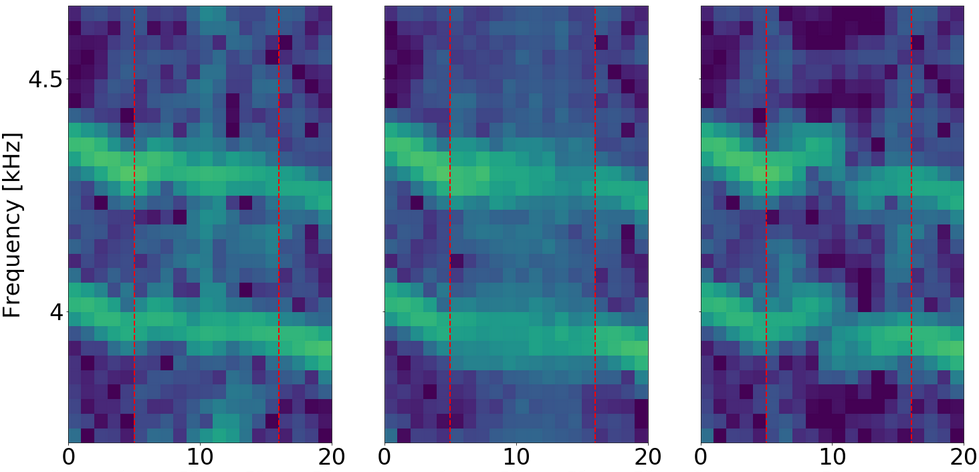}
\caption{Sections of magnitude spectrograms (in dB) of an exemplary signal reconstruction. Left: Original signal. Center: Reconstruction by the magnitude network. Right: Reconstruction by the LPC-based method. The magnitude network provided a 5~dB larger $\text{SNR}_{\text{MS}}$.}
\label{fig:spectrogram_zoom_shift}
\end{center}
\end{figure}

When looking more in the details of the reconstruction, both methods showed different characteristics: In Figure \ref{fig:spectrogram_zoom_shift} we show spectrograms of an instrument signal with frequency-modulated components. The LPC-based reconstruction shows a discontinuity in the middle of the gap instead of a steady transition. This is the consequence of the two extrapolations (forward and backwards), mixed in the middle of the gap. The magnitude network trained on the music learned how to represent frequency modulations and provides less artifacts in the reconstruction, which yielded a 5~dB larger $\text{SNR}_{\text{MS}}$. 

\begin{figure}[!th]
\begin{center}
\includegraphics[width=0.45\textwidth,height=.18\textheight]{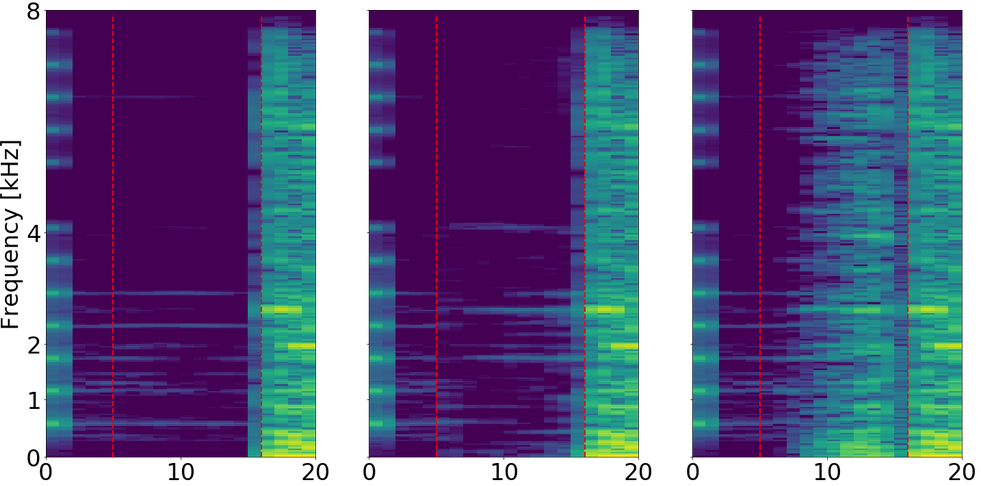}
\includegraphics[width=0.45\textwidth,height=.18\textheight]{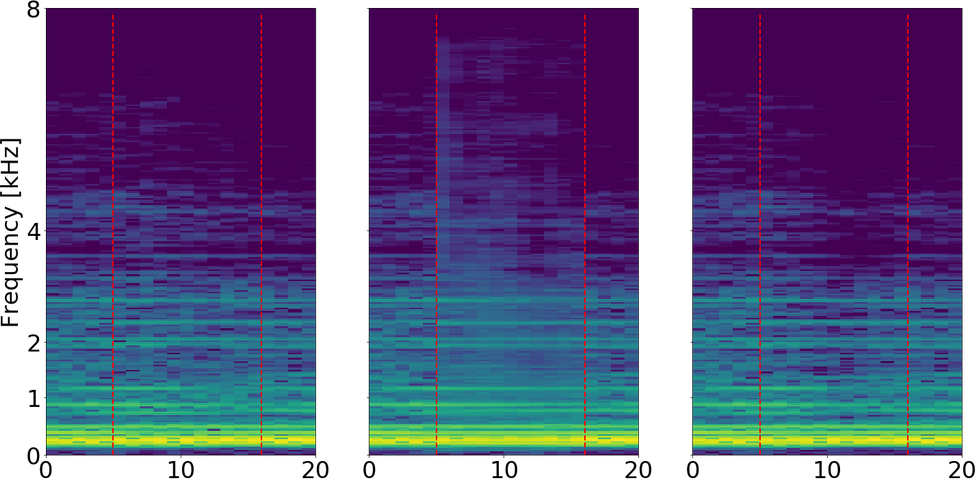}
\caption{Magnitude spectrograms (in dB) of exemplary signal reconstructions. Left: Original signal. Center: Reconstruction by the magnitude network. Right: Reconstruction by the LPC-based method. Top: Example with the magnitude network outperforming the reference by an $\text{SNR}_{\text{MS}}$ of 13~dB. Bottom: Example with the magnitude network underperforming the reference by an $\text{SNR}_{\text{MS}}$ of 9~dB.}
\label{fig:spectrogram_examples}
\end{center}
\end{figure}

Another interesting examples are shown in Figure \ref{fig:spectrogram_examples}. The top row shows an example in which the magnitude network outperformed the LPC-based method. In this case, the signal is comprised of steady harmonic tones in the left side context and a broadband sound in the right side context. While the LPC-based method extrapolated the broadband noise into the gap, the magnitude network was able to foresee the transition from the steady sounds to the broadband burst, yielding a prediction much closer to the original gap, with a 13~dB larger $\text{SNR}_{\text{MS}}$ than that from the LPC-based method.
	
On the other hand, the magnitude network not always outperformed the LPC-based method. The bottom row of Figure \ref{fig:spectrogram_examples} shows spectrograms of such an example. This signal had stable sounds in the gap, which were well-suited for an extrapolation, but rather complex to be perfectly reconstructed by the magnitude network. Thus, the LPC-based method outperformed the magnitude network yielding a 9~dB larger $\text{SNR}_{\text{MS}}$. 

The excellent performance of the LPC-based method reconstructing instruments can be explained by the assumptions behind the LPC well-fitting to the single-note instrument sounds. These sounds usually consist of harmonics stable on a short-time scale. LPC extrapolates these harmonics preserving the spectral envelope of the signal. Nevertheless, the magnitude network yielded an $\text{SNR}_{\text{MS}}$ of 22.0~dB, on average, demonstrating a good ability to reconstruct instrument sounds. 

When applied on music, the performance of both methods was much poorer, with our network performing slightly but statistically significantly better than the LPC-based method. The better performance of our network can be explained by its ability to adapt to transient sounds and modulations in frequencies, sound properties that the LPC-based method is not suited to handle.

\subsection{Effect of the gap duration}

The proposed network structure can be trained with different contexts and gap durations. For problems of varying gap duration, a network trained to the particular gap duration might appear optimal. However, training takes time, and it might be simpler to train a network to single gap duration and use it to reconstruct any shorter gap as well.

In order to test this idea, we introduced gaps of $48$~ms in our testing datasets. These gaps were then reconstructed by the network trained for $64$~ms gaps. As this network outputs, at reconstruction time, a solution for a gap of length $64$-ms, the $48$-ms gaps needs to be extended. We tested three types of the extension: $16$~ms forwards, $16$~ms backwards, and centered ($8$~ms forwards and $8$~ms backwards).

Table \ref{tab:SpectralDiv48} shows $\text{SNR}_{\text{MS}}$ obtained from reconstructions of the three types of gap extension, as averages over these extension types. Also, the corresponding $\text{SNR}_{\text{MS}}$ for the LPC-based method are shown. The results are similar to those obtained for larger gaps: for the instruments, the LPC-based method outperformed our network; for the music, our network outperformed the LPC-based method.

\begin{table}[!th]
\centering
\begin{tabular}{lcccc}
\hline
      & \multicolumn{2}{c}{Music} & \multicolumn{2}{c}{Instruments}\\ 
      & Ours & LPC & Ours & LPC  \\     
      \hline
  
Mean  & 8.0   & 6.9    & 21.6   & 33.2\\
Std   & 4.6   & 5.5    & 11.5 	& 20.1\\ \hline \\
\end{tabular}
\caption{$\text{SNR}_{\text{MS}}$ (in dB) of reconstructions of 48~ms gaps for the magnitude network and the LPC-based method.}
\label{tab:SpectralDiv48}
\vspace{-1em}
\end{table}

\section{Conclusions and Outlook}
\label{sec:Conclusion}
We proposed a convolutional neural network architecture working as a context encoder on TF coefficients. For the reconstruction of complex signals like music, that network was able to outperform the LPC-based reference method, in terms SNR calculated on magnitude spectrograms. However, LPC yielded better results when applied on more simple signals like instrument sounds. We have further shown that the proposed network was able to adapt to the particular pitches provided by the training material and that it can be applied to gaps shorter than the trained ones. In general, our results suggest that standard components and a moderately sized network can be applied to form audio-inpainting models, offering a number of angles for future improvement. 

Training a generative method directly on audio data requires vast size of datasets and large networks. To construct more compact models of moderately sized datasets, it is imperative to use efficient input audio features and an invertible feature representation at the output. Here, the STFT features, meant as a reasonable first choice, provided a decent performance. In the future, we expect more hearing-related features to provide even better reconstructions. In particular, an investigation of Audlet frames, i.e., invertible time-frequency systems adapted to perceptual frequency scales, \cite{necciari18}, as features for audio inpainting present intriguing opportunities.

Generally, better results can be expected for increased depth of the network and the available context. Unfortunately, our preliminary tests of simply increasing the network's depth led to minor improvements only. As it seems, a careful consideration of the building blocks of the model is required instead. Here, preferred architectures are those not relying on a predetermined target and input feature length, e.g., a recurrent network. Recent advances in generative networks will provide other interesting alternatives for analyzing and processing audio data as well. These approaches are yet to be fully explored.

Finally, music data can be highly complex and it is unreasonable to expect a single trained model to accurately inpaint a large number of musical styles and instruments at once. Thus, instead of training on a very general dataset, we expect significantly improved performance for more specialized networks that could be trained by restricting the training data to specific genres or instrumentation. Applied to a complex mixture and potentially preceded by a source-separation algorithm, the resulting models could be used jointly in a mixture-of-experts,~\cite{yuksel2012twenty}, approach.

\bibliographystyle{jaes}
\bibliography{context}

\begin{thebibliography}{48}
\newcommand{\enquote}[1]{``#1''}
\providecommand{\natexlab}[1]{#1}
\expandafter\ifx\csname urlstyle\endcsname\relax
  \providecommand{\doi}[1]{doi:\discretionary{}{}{}#1}\else
  \providecommand{\doi}{doi:\discretionary{}{}{}\begingroup
  \urlstyle{rm}\Url}\fi

\bibitem[{Pathak et~al.(2016)Pathak, Krahenbuhl, Donahue, Darrell, and
  Efros}]{Pathak2016}
Pathak, D., Krahenbuhl, P., Donahue, J., Darrell, T., and Efros, A.,
  \enquote{Context Encoders: Feature Learning by Inpainting,} 2016.

\bibitem[{Goodfellow et~al.(2016)Goodfellow, Bengio, and
  Courville}]{Goodfellow-et-al-2016}
Goodfellow, I., Bengio, Y., and Courville, A., \emph{Deep Learning}, MIT Press,
  2016, \url{http://www.deeplearningbook.org}.

\bibitem[{Pons et~al.(2017)Pons, Nieto, Prockup, Schmidt, Ehmann, and
  Serra}]{Pons2017}
Pons, J., Nieto, O., Prockup, M., Schmidt, E.~M., Ehmann, A.~F., and Serra, X.,
  \enquote{End-to-end learning for music audio tagging at scale,} \emph{CoRR},
  abs/1711.02520, 2017.

\bibitem[{{P}ortnoff(1976)}]{po76}
{P}ortnoff, M., \enquote{{I}mplementation of the digital phase vocoder using
  the fast fourier transform,} \emph{{I}{E}{E}{E} {T}rans. {A}coust. {S}peech
  {S}ignal {P}rocess.}, 24(3), pp. 243--248, 1976.

\bibitem[{{G}r{\"o}chenig(2001)}]{gr01}
{G}r{\"o}chenig, K., \emph{{F}oundations of {T}ime-{F}requency {A}nalysis},
  {A}ppl. {N}umer. {H}armon. {A}nal., {B}irkh{\"a}user, 2001.

\bibitem[{Griffin and Lim(1984)}]{griffin1984signal}
Griffin, D. and Lim, J., \enquote{Signal estimation from modified short-time
  Fourier transform,} \emph{IEEE Transactions on Acoustics, Speech and Signal
  Processing}, 32(2), pp. 236--243, 1984.

\bibitem[{Perraudin et~al.(2013)Perraudin, Balazs, and
  S{\o}ndergaard}]{perraudin2013fast}
Perraudin, N., Balazs, P., and S{\o}ndergaard, P.~L., \enquote{A fast
  Griffin-Lim algorithm,} in \emph{Applications of Signal Processing to Audio
  and Acoustics (WASPAA), 2013 IEEE Workshop on}, pp. 1--4, IEEE, 2013.

\bibitem[{Pr{\r{u}}{\v{s}}a et~al.(2017)Pr{\r{u}}{\v{s}}a, Balazs, and
  S{\o}ndergaard}]{pruuvsa2017noniterative}
Pr{\r{u}}{\v{s}}a, Z., Balazs, P., and S{\o}ndergaard, P., \enquote{A
  noniterative method for reconstruction of phase from STFT magnitude,}
  \emph{IEEE/ACM Transactions on Audio, Speech and Language Processing}, 25(5),
  pp. 1154--1164, 2017.

\bibitem[{Schl{\"u}ter(2017)}]{schlueter2017_phd}
Schl{\"u}ter, J., \emph{{Deep Learning for Event Detection, Sequence Labelling
  and Similarity Estimation in Music Signals}}, Ph.D. thesis, Johannes Kepler
  University Linz, Austria, 2017.

\bibitem[{Kingma and Welling(2013)}]{KingmaW13}
Kingma, D. and Welling, M., \enquote{Auto-Encoding Variational Bayes.}
  \emph{CoRR}, abs/1312.6114, 2013.

\bibitem[{Goodfellow et~al.(2014)Goodfellow, Pouget-Abadie, Mirza, Xu,
  Warde-Farley, Ozair, Courville, and Bengio}]{GoodfellowGAN2014}
Goodfellow, I., Pouget-Abadie, J., Mirza, M., Xu, B., Warde-Farley, D., Ozair,
  S., Courville, A., and Bengio, Y., \enquote{Generative adversarial nets,} in
  \emph{Advances in neural information processing systems}, pp. 2672--2680,
  2014.

\bibitem[{Donahue et~al.(2018)Donahue, McAuley, and Puckette}]{Donahue2018}
Donahue, C., McAuley, J., and Puckette, M., \enquote{{Synthesizing Audio with
  Generative Adversarial Networks},} \emph{ArXiv e-prints}, 2018.

\bibitem[{Mehri et~al.(2016)Mehri, Kumar, Gulrajani, Kumar, Jain, Sotelo,
  Courville, and Bengio}]{Mehri2016}
Mehri, S., Kumar, K., Gulrajani, I., Kumar, R., Jain, S., Sotelo, J.,
  Courville, A., and Bengio, Y., \enquote{SampleRNN: An Unconditional
  End-to-End Neural Audio Generation Model,} \emph{CoRR}, abs/1612.07837, 2016.

\bibitem[{van~den Oord et~al.(2016)van~den Oord, Dieleman, Zen, Simonyan,
  Vinyals, Graves, Kalchbrenner, Senior, and Kavukcuoglu}]{Wavenet2016}
van~den Oord, A., Dieleman, S., Zen, H., Simonyan, K., Vinyals, O., Graves, A.,
  Kalchbrenner, N., Senior, A., and Kavukcuoglu, K., \enquote{WaveNet: {A}
  Generative Model for Raw Audio,} \emph{CoRR}, abs/1609.03499, 2016.

\bibitem[{Wang et~al.(2017)Wang, Skerry{-}Ryan, Stanton, Wu, Weiss, Jaitly,
  Yang, Xiao, Chen, Bengio, Le, Agiomyrgiannakis, Clark, and
  Saurous}]{tacotron}
Wang, Y., Skerry{-}Ryan, R., Stanton, D., Wu, Y., Weiss, R., Jaitly, N., Yang,
  Z., Xiao, Y., Chen, Z., Bengio, S., Le, Q., Agiomyrgiannakis, Y., Clark, R.,
  and Saurous, R., \enquote{Tacotron: {A} Fully End-to-End Text-To-Speech
  Synthesis Model,} \emph{CoRR}, abs/1703.10135, 2017.

\bibitem[{Shen et~al.(2017)Shen, Pang, Weiss, Schuster, Jaitly, Yang, Chen,
  Zhang, Wang, Skerry{-}Ryan, Saurous, Agiomyrgiannakis, and
  Wu}]{Tacotron2_2017}
Shen, J., Pang, R., Weiss, R., Schuster, M., Jaitly, N., Yang, Z., Chen, Z.,
  Zhang, Y., Wang, Y., Skerry{-}Ryan, R., Saurous, R., Agiomyrgiannakis, Y.,
  and Wu, Y., \enquote{Natural {TTS} Synthesis by Conditioning WaveNet on Mel
  Spectrogram Predictions,} \emph{CoRR}, abs/1712.05884, 2017.

\bibitem[{Lee and Chang(2016)}]{Lee2016PLC}
Lee, B.-K. and Chang, J.-H., \enquote{Packet Loss Concealment Based on Deep
  Neural Networks for Digital Speech Transmission,} \emph{IEEE/ACM Trans.
  Audio, Speech and Lang. Proc.}, 24(2), pp. 378--387, 2016, ISSN 2329-9290,
  \doi{10.1109/TASLP.2015.2509780}.

\bibitem[{Dieleman et~al.(2018)Dieleman, Oord, and
  Simonyan}]{dieleman2018challenge}
Dieleman, S., Oord, A. v.~d., and Simonyan, K., \enquote{The challenge of
  realistic music generation: modelling raw audio at scale,} \emph{arXiv
  preprint arXiv:1806.10474}, 2018.

\bibitem[{Boulanger-Lewandowski et~al.(2012)Boulanger-Lewandowski, Bengio, and
  Vincent}]{Boulanger2012ModelingTD}
Boulanger-Lewandowski, N., Bengio, Y., and Vincent, P., \enquote{Modeling
  Temporal Dependencies in High-Dimensional Sequences: Application to
  Polyphonic Music Generation and Transcription,} in \emph{ICML}, 2012.

\bibitem[{Blaauw and Bonada(2017)}]{Blaauw2017}
Blaauw, M. and Bonada, J., \enquote{A Neural Parametric Singing Synthesizer,}
  \emph{CoRR}, abs/1704.03809, 2017.

\bibitem[{Adler et~al.(2011)Adler, Emiya, Jafari, Elad, Gribonval, and
  Plumbley}]{Adler2011}
Adler, A., Emiya, V., Jafari, M., Elad, M., Gribonval, R., and Plumbley, M.,
  \enquote{A constrained matching pursuit approach to audio declipping,} in
  \emph{{IEEE} International Conference on Acoustics, Speech and Signal
  Processing ({ICASSP})}, 2011, \doi{10.1109/icassp.2011.5946407}.

\bibitem[{Adler et~al.(2012)Adler, Emiya, Jafari, Elad, Gribonval, and
  Plumbley}]{Adler2012}
Adler, A., Emiya, V., Jafari, M.~G., Elad, M., Gribonval, R., and Plumbley,
  M.~D., \enquote{Audio Inpainting,} \emph{IEEE Transactions on Audio, Speech
  and Language Processing}, 20(3), pp. 922--932, 2012,
  \doi{10.1109/TASL.2011.2168211}.

\bibitem[{Toumi and Emiya(2018)}]{toumi:hal-01680669}
Toumi, I. and Emiya, V., \enquote{{Sparse non-local similarity modeling for
  audio inpainting},} in \emph{{ICASSP - IEEE International Conference on
  Acoustics, Speech and Signal Processing}}, Calgary, Canada, 2018.

\bibitem[{Siedenburg et~al.(2013)Siedenburg, D{\"o}rfler, and
  Kowalski}]{Siedenburg2013AudioIW}
Siedenburg, K., D{\"o}rfler, M., and Kowalski, M., \enquote{Audio inpainting
  with social sparsity,} \emph{SPARS (Signal Processing with Adaptive Sparse
  Structured Representations)}, 2013.

\bibitem[{Lieb and Stark(2018)}]{Lieb2018}
Lieb, F. and Stark, H.-G., \enquote{Audio inpainting: Evaluation of
  time-frequency representations and structured sparsity approaches,}
  \emph{Signal Processing}, 153, pp. 291--299, 2018.

\bibitem[{Bahat et~al.(2015)Bahat, Schechner, and Elad}]{Bahat2015}
Bahat, Y., Schechner, Y., and Elad, M., \enquote{Self-content-based audio
  inpainting,} \emph{Signal Processing}, 111, pp. 61--72, 2015,
  \doi{10.1016/j.sigpro.2014.11.023}.

\bibitem[{Perraudin et~al.(2018)Perraudin, Holighaus, Majdak, and
  Balazs}]{Perraudin2016}
Perraudin, N., Holighaus, N., Majdak, P., and Balazs, P., \enquote{Inpainting
  of long audio segments with similarity graphs,} \emph{IEEE/ACM Transactions
  on Audio, Speech and Language Processing}, PP(99), pp. 1--1, 2018, ISSN
  2329-9290, \doi{10.1109/TASLP.2018.2809864}.

\bibitem[{Etter(1996)}]{Etter1996}
Etter, W., \enquote{Restoration of a discrete-time signal segment by
  interpolation based on the left-sided and right-sided autoregressive
  parameters,} \emph{{IEEE} Transactions on Signal Processing}, 44(5), pp.
  1124--1135, 1996, \doi{10.1109/78.502326}.

\bibitem[{Kauppinen and Roth(2002)}]{kauppinen2002audio}
Kauppinen, I. and Roth, K., \enquote{Audio signal extrapolation--theory and
  applications,} in \emph{Proc. DAFx}, pp. 105--110, 2002.

\bibitem[{Srivastava et~al.(2014)Srivastava, Hinton, Krizhevsky, Sutskever, and
  Salakhutdinov}]{srivastava2014dropout}
Srivastava, N., Hinton, G., Krizhevsky, A., Sutskever, I., and Salakhutdinov,
  R., \enquote{Dropout: a simple way to prevent neural networks from
  overfitting,} \emph{The Journal of Machine Learning Research}, 15(1), pp.
  1929--1958, 2014.

\bibitem[{Abadi et~al.(2015)Abadi, Agarwal, Barham, Brevdo, Chen, Citro,
  Corrado, Davis, Dean, Devin, Ghemawat, Goodfellow, Harp, Irving, Isard, Jia,
  Jozefowicz, Kaiser, Kudlur, Levenberg, Man\'{e}, Monga, Moore, Murray, Olah,
  Schuster, Shlens, Steiner, Sutskever, Talwar, Tucker, Vanhoucke, Vasudevan,
  Vi\'{e}gas, Vinyals, Warden, Wattenberg, Wicke, Yu, and
  Zheng}]{tensorflow2015-whitepaper}
Abadi, M., Agarwal, A., Barham, P., Brevdo, E., Chen, Z., Citro, C., Corrado,
  G., Davis, A., Dean, J., Devin, M., Ghemawat, S., Goodfellow, I., Harp, A.,
  Irving, G., Isard, M., Jia, Y., Jozefowicz, R., Kaiser, L., Kudlur, M.,
  Levenberg, J., Man\'{e}, D., Monga, R., Moore, S., Murray, D., Olah, C.,
  Schuster, M., Shlens, J., Steiner, B., Sutskever, I., Talwar, K., Tucker, P.,
  Vanhoucke, V., Vasudevan, V., Vi\'{e}gas, F., Vinyals, O., Warden, P.,
  Wattenberg, M., Wicke, M., Yu, Y., and Zheng, X., \enquote{{TensorFlow}:
  Large-Scale Machine Learning on Heterogeneous Systems,} 2015, software
  available from tensorflow.org.

\bibitem[{Kingma and Ba()}]{Kingma2014}
Kingma, D. and Ba, J., \enquote{Adam: A Method for Stochastic Optimization,}
\emph{online:} arxiv.org/pdf/1412.6980v9.
  
  
\bibitem[{Ramachandran et~al.()Ramachandran, Zoph, and Le}]{Ramachandran2017}
Ramachandran, P., Zoph, B., and Le, Q., \enquote{Searching for Activation
  Functions,} \emph{online:} arxiv.org/pdf/1710.05941v2.

\bibitem[{Ioffe and Szegedy(2015)}]{Ioffe2015}
Ioffe, S. and Szegedy, C., \enquote{Batch Normalization: Accelerating Deep
  Network Training by Reducing Internal Covariate Shift,} \emph{CoRR},
  abs/1502.03167, 2015.

\bibitem[{Pr\r{u}\v{s}a and S{\o}ndergaard(2016)}]{ltfatnote043}
Pr\r{u}\v{s}a, Z. and S{\o}ndergaard, P.~L., \enquote{{Real-Time Spectrogram
  Inversion Using Phase Gradient Heap Integration},} in \emph{Proc. Int. Conf.
  Digital Audio Effects (DAFx-16)}, pp. 17--21, 2016.

\bibitem[{Pr\r{u}\v{s}a(2017)}]{ltfatnote045}
Pr\r{u}\v{s}a, Z., \enquote{{The Phase Retrieval Toolbox},} in \emph{{AES}
  International Conference On Semantic Audio}, Erlangen, Germany, 2017.

\bibitem[{Zhao et~al.(2017)Zhao, Gallo, Frosio, and Kautz}]{ZhaoLoss2017}
Zhao, H., Gallo, O., Frosio, I., and Kautz, J., \enquote{Loss Functions for
  Image Restoration With Neural Networks,} \emph{IEEE Transactions on
  Computational Imaging}, 3(1), pp. 47--57, 2017, ISSN 2333-9403,
  \doi{10.1109/TCI.2016.2644865}.

\bibitem[{Krogh and Hertz(1992)}]{Krogh92asimple}
Krogh, A. and Hertz, J., \enquote{A Simple Weight Decay Can Improve
  Generalization,} in \emph{Advances in neural information processing systems
  4}, pp. 950--957, Morgan Kaufmann, 1992.

\bibitem[{Engel et~al.(2017)Engel, Resnick, Roberts, Dieleman, Eck, Simonyan,
  and Norouzi}]{nsynth2017}
Engel, J., Resnick, C., Roberts, A., Dieleman, S., Eck, D., Simonyan, K., and
  Norouzi, M., \enquote{Neural Audio Synthesis of Musical Notes with WaveNet
  Autoencoders,} 2017.

\bibitem[{Defferrard et~al.(2017)Defferrard, Benzi, Vandergheynst, and
  Bresson}]{fma2017}
Defferrard, M., Benzi, K., Vandergheynst, P., and Bresson, X., \enquote{FMA: A
  Dataset for Music Analysis,} in \emph{18th International Society for Music
  Information Retrieval Conference}, 2017.

\bibitem[{Sturmel and Daudet(2011)}]{sturmel2011signal}
Sturmel, N. and Daudet, L., \enquote{Signal reconstruction from STFT magnitude:
  A state of the art,} in \emph{International conference on digital audio
  effects (DAFx)}, pp. 375--386, 2011.

\bibitem[{Tremain(1982)}]{Tremain1982}
Tremain, T.~E., \enquote{The Government Standard Linear Predictive Coding
  Algorithm: LPC-10,} \emph{Speech Technology}, pp. 40--49, 1982.

\bibitem[{Rajman and Pallota(2007)}]{rajman2007speech}
Rajman, M. and Pallota, V., \emph{Speech and language engineering}, EPFL Press,
  2007.

\bibitem[{Kauppinen et~al.(2001)Kauppinen, Kauppinen, and
  Saarinen}]{kauppinen2001method}
Kauppinen, I., Kauppinen, J., and Saarinen, P., \enquote{A method for long
  extrapolation of audio signals,} \emph{Journal of the Audio Engineering
  Society}, 49(12), pp. 1167--1180, 2001.

\bibitem[{Burg(1967)}]{Burg1967}
Burg, J.~P., \enquote{Maximum entropy spectral analysis,} \emph{37th Annual
  International Meeting, Soc. of Explor. Geophys., Oklahoma City}, 1967.

\bibitem[{Kauppinen and Kauppinen(2002)}]{kauppinen2002reconstruction}
Kauppinen, I. and Kauppinen, J., \enquote{Reconstruction method for missing or
  damaged long portions in audio signal,} \emph{Journal of the Audio
  Engineering Society}, 50(7/8), pp. 594--602, 2002.

\bibitem[{Necciari et~al.(2018)Necciari, Holighaus, Balazs, Průša, Majdak,
  and Derrien}]{necciari18}
Necciari, T., Holighaus, N., Balazs, P., Průša, Z., Majdak, P., and Derrien,
  O., \enquote{Audlet Filter Banks: A Versatile Analysis/Synthesis Framework
  Using Auditory Frequency Scales,} \emph{Applied Sciences}, 8(1:96), 2018.

\bibitem[{Yuksel et~al.(2012)Yuksel, Wilson, and Gader}]{yuksel2012twenty}
Yuksel, S.~E., Wilson, J.~N., and Gader, P.~D., \enquote{Twenty years of
  mixture of experts,} \emph{IEEE transactions on neural networks and learning
  systems}, 23(8), pp. 1177--1193, 2012.

\end{thebibliography}

\end{document}